# Multifocus microscopy with optically sectioned axial superresolution

Florian Ströhl*1, Daniel Henry Hansen1, Mireia Nager Grifo2, Åsa Birna Birgisdottir2,3

1 Department of Physics and Technology, UiT The Arctic University of Norway, Tromsø, Norway

2 Division of Cardiothoracic and Respiratory Medicine, University Hospital of North Norway, Tromsø, Norway

3 Department of Clinical Medicine, UiT The Arctic University of Norway, Tromsø, Norway

* florian.strohl@uit.no

## Abstract

Multifocus microscopy enables recording of entire volumes in a single camera exposure. In dense samples, multifocus microscopy is severely hampered by background haze. Here, we introduce a scalable multifocus method that incorporates optical sectioning and offers axial superresolution capabilities. In our method, a dithered oblique light-sheet scans the sample volume during a single exposure, while generated fluorescence is linearised onto the camera with a multifocus optical element. A synchronised rolling shutter readout realised optical sectioning. We describe the technique theoretically and verify its optical sectioning and superresolution capabilities. We demonstrate a prototype system with a multifocus beam splitter cascade and record monolayers of endothelial cells at 35 volumes per second. We furthermore image uncleared engineered human heart tissue and visualise the distribution of mitochondria at axial superresolution. Our method manages to capture sub-diffraction sized mitochondria-derived vesicles up to 30 µm deep into the tissue.

# Introduction

Capturing fast 3D processes on the subcellular level is a recurring challenge in fluorescence microscopy [1]. Most imaging modalities, like confocal, spinning disk or light-sheet microscopy perform sequential recording of either points or planes. The required scanning has two downsides: sample movement during acquisition can lead to artefacts and mechanically changing focus to different planes can perturb the sample itself. The inclusion of remote-focusing [2]–[6] can mitigate the latter, but still requires sequential acquisition of many individual frames in addition to time-costly image volume de-shearing. Furthermore, illumination light in confocal and spinning disk microscopy extends beyond the imaged voxel or plane, thus contributing heavily to photobleaching of fluorophores and light-induced sample damage. Computational widefield approaches based on deconvolution and neural networks that promise 3D capabilities [7]–[10] share this drawback.

For optimal use of the illumination volume and to maximise imaging speed, the acquisition of entire focal stacks in each camera frame is an elegant solution. Multifocus imaging methods exist [11]–[13], but current implementations generally do not enable optical sectioning and therefore have poor axial resolution. The existing variants can be broadly grouped by the employed multi-plane optical element into *reflective*, *refractive*, and *diffractive* types. Concatenations of beam-splitters [13], [14], multi-plane prisms [15] and similar stacked-partial-reflection approaches [16], [17] divide the nominal image plane into several sub-planes that reach different locations of the camera sensor with different optical path lengths, which relate to different focal planes. Defocusing achieved in such a way induces spherical aberrations that increase with focal plane shift. If aberrations are left uncorrected, imaging is restricted to shallow volumes of a few micrometres depth [13]; If aberrations are to be corrected, a separate set of corrective optics for each focal plane is required [17], which hampers scalability.

The most prominent *refractive* multifocus method is light-field microscopy [11]. It uses microlens arrays, which allow recording of information on both location and direction of emission light (together known as light field), which permits computation of the 3D sample distribution. The light field dataset contains enough information to mitigate depth-induced optical aberrations during the image reconstruction process. On the downside, light-field microscopy suffers an inherent trade-off between resolution and field of view, non-isotropic resolution across z planes, and reconstruction artefacts in the presence of stray light [18].

Multifocus microscopy based on warped gratings [12] splits an image into diffraction orders with order-dependent defocus. Spherical aberrations due to defocus are countered by designed grating-induced spherical aberrations of opposite sign. All image planes are thus free of *monochromatic* aberrations. Spectrally broad fluorophores exhibit strong chromatic aberrations though, which need to be corrected using additional gratings and prisms [19]. When implemented in such a way, this method is capable of imaging dozens of planes spanning tens of micrometres in depth [20].

Ideally, multifocus microscopy should provide confocal resolution and background rejection, be as fast, versatile, and gentle as light-sheet microscopy, while still allow conventional sample mounting. In the following, we demonstrate how dithered oblique plane light-sheet illumination can be combined with multifocus microscopy to realise optically sectioned single-shot volume imaging.

## Results

Consider a light-sheet sweeping through the sample volume as depicted in Figure 1. At each instant in time, a multifocus imager can linearise the illuminated plane onto a single row on a camera sensor. Once a full sweep is complete, the entire volume is mapped onto the plane of the camera. Depending on the camera's read-out mode, two different cases can be distinguished. Firstly, if the camera's shutter is open for the full duration of the sweep, the recorded image is indistinguishable from an image taken under widefield illumination and is hence corrupted by background haze. The respective optical transfer function (OTF) is that of widefield microscopy and lacks spatial frequencies along the optical axis. In the second case, each sensor line is read individually and synchronously with the light-sheet sweep. Therefore, the camera itself realises a pinhole effect and the effective OTF is governed by a convolution between illumination and detection OTFs. A light-sheet propagating *along the optical axis* thus provides optical sectioning at the widefield resolution limit, while a light-sheet swept at an *oblique angle* results in an image with axial superresolution. Note that the multifocus imager needs to be aligned differently under oblique illumination to match the camera's rolling shutter. In practice, this can be achieved by tilting the multiplane optical element or adjustment of its constituting parts in the case of a beam splitter cascade. Let us abbreviate this approach as SOLIS (**s**canned **o**blique **l**ight-sheet **i**nstant-volume **s**ectioning).

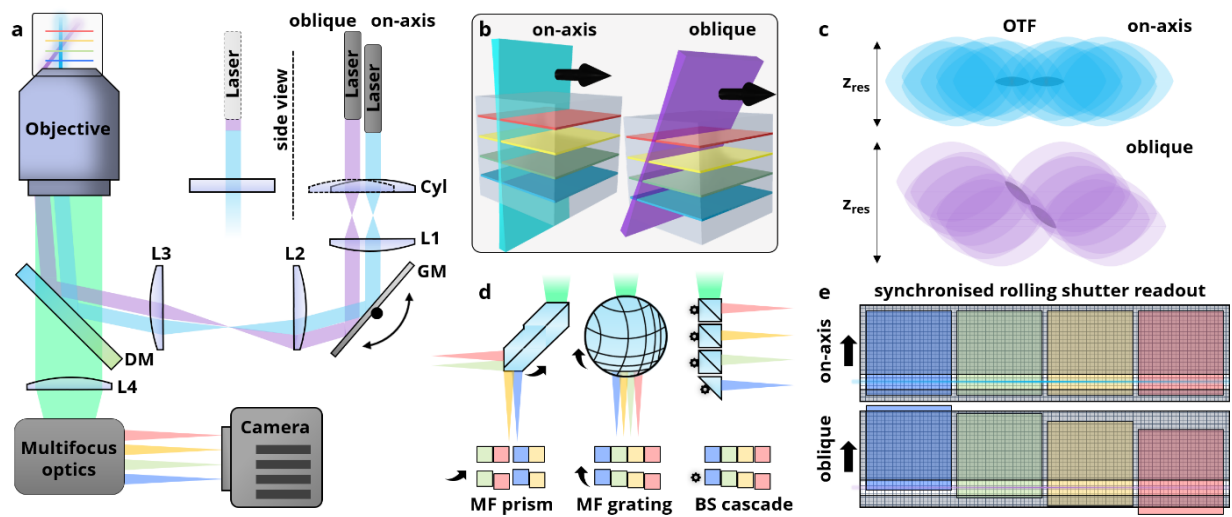

*Figure 1: Concept of instant volume imaging. (a) Depicted in light blue is the light-path for an on-axis light-sheet. Light purple shows the light-path for an oblique light-sheet. Fluorescence from the imaging volume is collected episcopically and relayed via multifocus optics onto a camera chip. (b) Schematic of the illumination geometries with swept on-axis and oblique light-sheet. Various focal planes are highlighted with colour coding. (c). The effective optical transfer function (OTF) is a convolution between light-sheet and detection OTFs. Both light-sheet geometries fill the missing cone and oblique illumination furthermore improves the achievable axial resolution. (d) Different types of multifocus elements (multifocus prisms* [15]*, multifocus gratings* [19]*, beam splitter cascades* [14]*), which generate displaced images of*

*corresponding focal planes on a camera chip. Adjustment of the multifocus elements permits linearisation of oblique light-sheets. (e) Synchronised rolling-shutter and light-sheet sweep during a single frame. Obliquely illuminated planes are displaced for alignment with the rolling shutter.*

To gauge the performance of SOLIS, we simulated 3D imaging in widefield microscopy with a scanned light-sheet, SOLIS with an on-axis propagating light-sheet, and with an oblique light-sheet. Referring to results displayed in Figure 2a-c, we find an effective elimination of background haze as expected. Optical sectioning performance and achievable resolution are determined from the respective optical transfer functions and here we find good agreement with theory in terms of expected resolution cut-offs in all dimensions. In particular, Table 1 lists a comparison of theoretical resolution values with measurements from the simulations at a signal-to-noise ratio (SNR) of 50. Note that the resolution cut-offs are in practice limited by signal-to-noise ratio. Figure 2d-g shows the strength of the optical transfer functions in a logarithmic scale sliced in the $k_y = 0$ plane.

*Table 1: Comparison of resolution measurements of 3D simulations with 50 SNR compared to theoretical values based on an XY geometrical analysis.*

|  | Δx [μm] | | Δy [μm] | | Δz [μm] | |
| --- | --- | --- | --- | --- | --- | --- |
|  | simulation | theory | simulation | theory | simulation | theory |
| **Widefield** | 0.21 | 0.20 | 0.21 | 0.20 | 0.53 | 0.53 |
| **On-axis** | 0.17 | 0.15 | 0.21 | 0.20 | 0.53 | 0.49 |
| **Oblique** | 0.19 | 0.18 | 0.21 | 0.20 | 0.37 | 0.33 |

Apart from a filled *missing cone* [21] and an almost doubled axial resolution cut-off, we also find a 5 times stronger OTF support in the case of SOLIS at higher axial spatial frequencies compared to widefield imaging. This is especially the case for oblique illumination, which suggests better performance at low-light conditions. The resolution gain in practice is dependent on the available signal-to-noise ratio, whereby a higher noise floor renders the periphery of the OTF challenging to use and thus can impact performance in practical scenarios even stronger than the absolute resolution cut-off. Note that denoising [22] and deconvolution [23] approaches exist that may rescue some degraded image information and can push the achievable resolution closer to the theoretical cut-off.

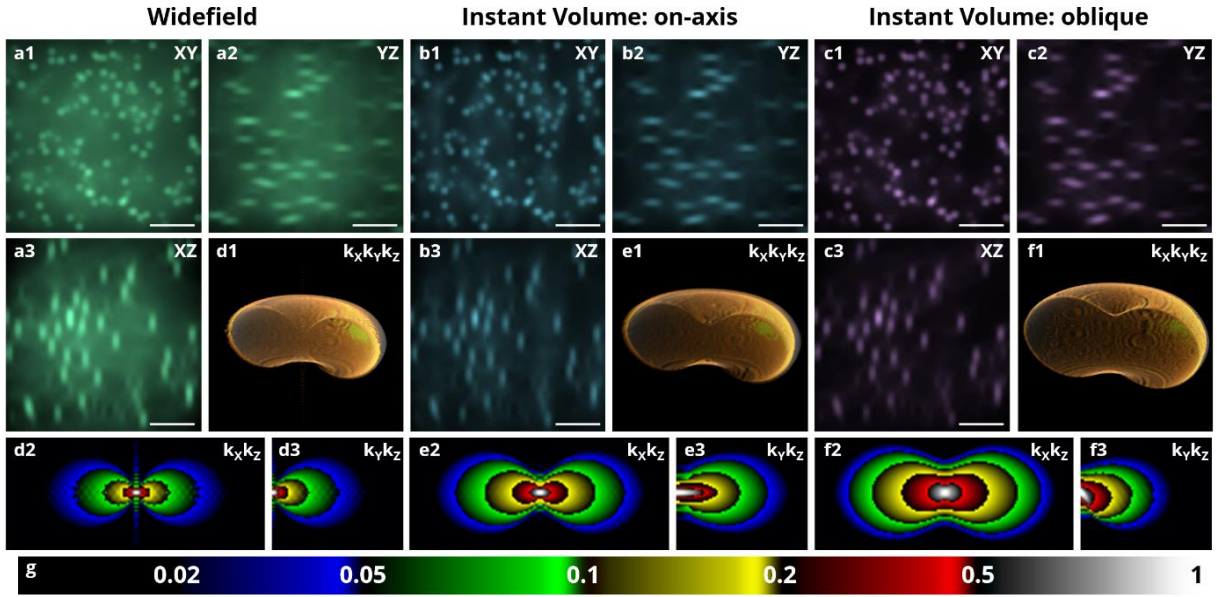

Figure 2: Simulation of SOLIS. (a-c) A cubic volume of 6.4 µm side length containing randomly distributed point emitters ($\lambda$ = 550 nm) was simulated with widefield and SOLIS imaging models. In b, the light-sheet is propagating along the optic axis; in c the light-sheet is oblique. Panel numbers correspond to XY, YZ, and XZ sections through the volume. Scale bars are 1.5 µm. (d-f) Simulations were repeated for a single emitter and 3D Fourier transformed. d1, e1, f1 show renderings of the outermost optical transfer function supports.

We proceeded by construction a SOLIS microscope as depicted in Figure 1a, equipped with a 3-plane beam splitter cascade for multifocus imaging (referred to as 3x1 splitter as 3 planes are sent onto 1 camera). We estimated the achievable spatial resolution twofold. First, we imaged 200 nm diameter fluorescent beads (Figure 3d) and fitted Gaussian functions through line profiles (Figure 1e). In the nominal focus plane, we find full width at half maximum (FWHM) values of 259 nm along the sheet, 239 nm across the sheet and 636 nm axially. The 200 nm beads proved to be more photostable than smaller beads but caused a deviation of the real PSF size from the measured one. Taking the size of the beads into account (see Methods), the resolution values correspond to 100 nm beads with FWHM of 206 nm along the sheet, 189 nm across the sheet, and 505 nm axially. We thus find very good agreement with the simulation results for SOLIS in the lateral dimensions. Bead size corrected widefield measurements are 215 nm, 224 nm, and 599 nm and we thus find an increase of SOLIS over widefield microscopy in the axial direction of nearly 100 nm. As is apparent in the XZ views of Figure 3d, the measured PSFs were affected by spherical aberrations, which are prone to limit performance, and can explain the discrepancy in theoretical and measured axial resolution gains on these beads. As alternative measure, we used phase decorrelation analysis [24] to gauge the resolution of actin stained bovine pulmonary artery endothelial (BPAE) cells and find a resolution down to 240 nm without any post-processing.

Using the same BPAE cells, we also demonstrate the efficiency of out-of-focus light rejection. Here, we used a 40x objective, which resulted in a spacing between focal planes of -5.6 µm and +5.4 µm respectively for above and below the nominal focus. Note that the plane separation is governed by the magnification of the microscope and the geometry of the beam splitter cascade (see Methods). As visualised in Figure 3a and b, SOLIS manages to remove out of focus light very effectively, resulting in clean optical sections. This can be done at high speed as exemplified in Figure 3c, where we imaged BPAE cells at a volumetric frame rate of 35 Hz while moving the stage at a speed of 35 µm/s. The resulting images are free from noticeable motion blur, yet optical sectioning is fully achieved in the case of SOLIS (see Visualisation 1).

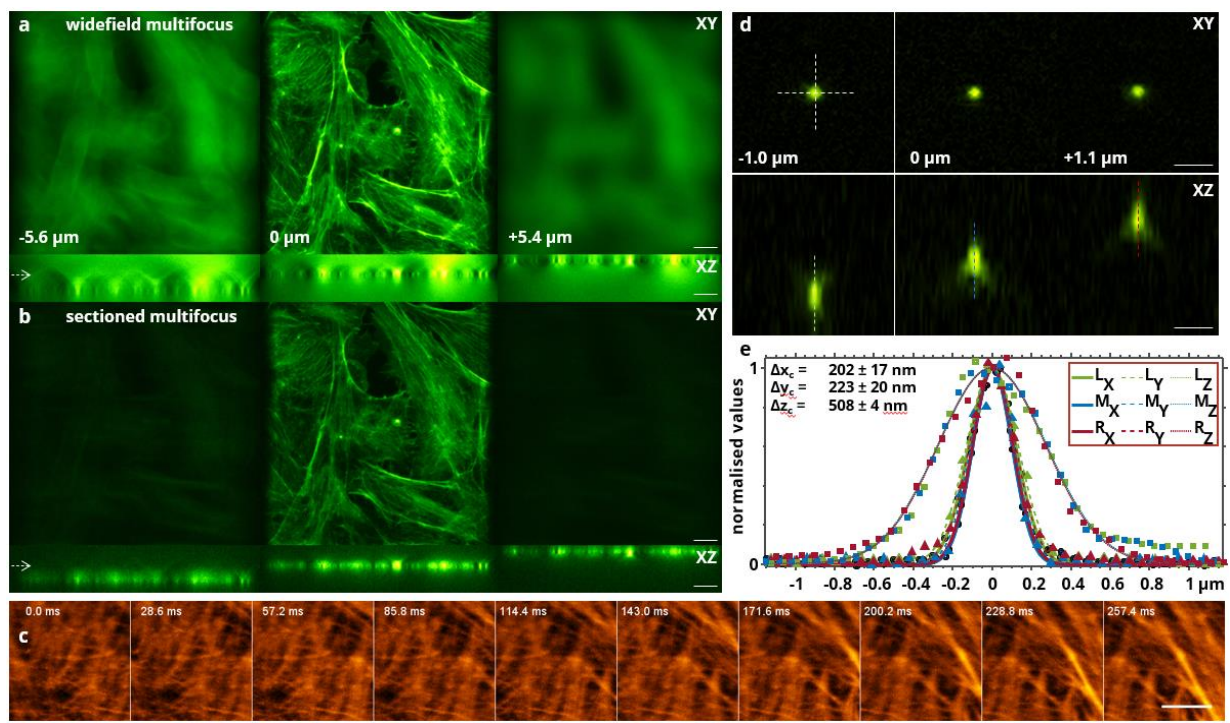

*Figure 3: SOLIS imaging. (a) Fixed Bovine Pulmonary Artery Endothelial (BPAE) cells with Alexa Fluor™ 488 Phalloidin labelled actin imaged on a 3x1 beam-splitter cascade multifocus microscope without and (b) with SOLIS using a 0.95 NA 40x air objective. Arrows in the XZ views refer to the displayed XY slice. (c) 10 frames of BPAE cells imaged at 35 Hz with the stage moving at 35 µm/s. A 1.35 NA 100x objective was used. (d) 200 nm diameter beads imaged with SOLIS using a 1.35 NA 100x silicone immersion objective. (e) Line profiles and Gaussian fits of PSFs from (c) in X, Y, Z. Reported resolution values are averaged between focal planes and corrected for bead size (see Methods). All scale bars are 1 µm.*

BPAE cells are thin and hence did generally not extend into off-focus planes. To fully demonstrate the volumetric imaging capabilities of SOLIS, we therefore imaged a more challenging three-dimensional sample: engineered human heart tissue (EHT). This type of tissue has high clinical relevance as conventional cell cultures of heart cells like cardiomyocytes do generally not fully mature [25], while ethical reasons limit the availability of primary human heart tissue. In contrast, EHTs are grown from induced pluripotent stem cells and cultured on special racks that permit synchronisation of the cells' contractions. After few weeks of culturing, a slab of beating heart tissue develops, which displays all crucial hallmarks of adult cardiac muscle. The tissue itself is dense and highly scattering, which complicates imaging of details in techniques without dedicated background rejection. In our tissue, we labelled mitochondria with TOM20 and imaged them with conventional multifocus microscopy as well as with SOLIS.

SOLIS' instant volume performance is strikingly displayed in Figure 4a,b and Visualisation 2, where the 3D distributions of mitochondria in cardiomyocytes within the tissue are recorded in a single camera exposure and individual mitochondria can be attributed to various z-positions with ease. In contrast, conventional multifocus microscopy is hampered severely by strong background haze. To test the limits of our technique, we performed multifocus imaging down to 75 µm into the tissue. Examples are shown in the inlays of Figure 4c,d and Visualisation 3. SOLIS generally manages to visualise individual mitochondria with a resolution of 0.3 µm to 0.4 µm down to 15 µm into the tissue based on decorrelation analysis [24]. Conventional multifocus imaging provides 0.5 µm to 0.6 µm resolution (decorrelation analysis) in this sample, presumable due to the severe background. Notably, SOLIS can resolve mitochondria-derived vesicles (MDVs) tens of micrometres deep into the tissue with measured sizes of around 260 nm lateral and 520 nm axial. Widefield multifocus microscopy is challenged in this environment and – if detectable at all - depicts MDVs with sizes of around 510 nm lateral and 700 nm axial (FWHM). Beyond 50 µm depth, the resolution drops to around 1 µm for SOLIS and below 2 µm for conventional imaging.

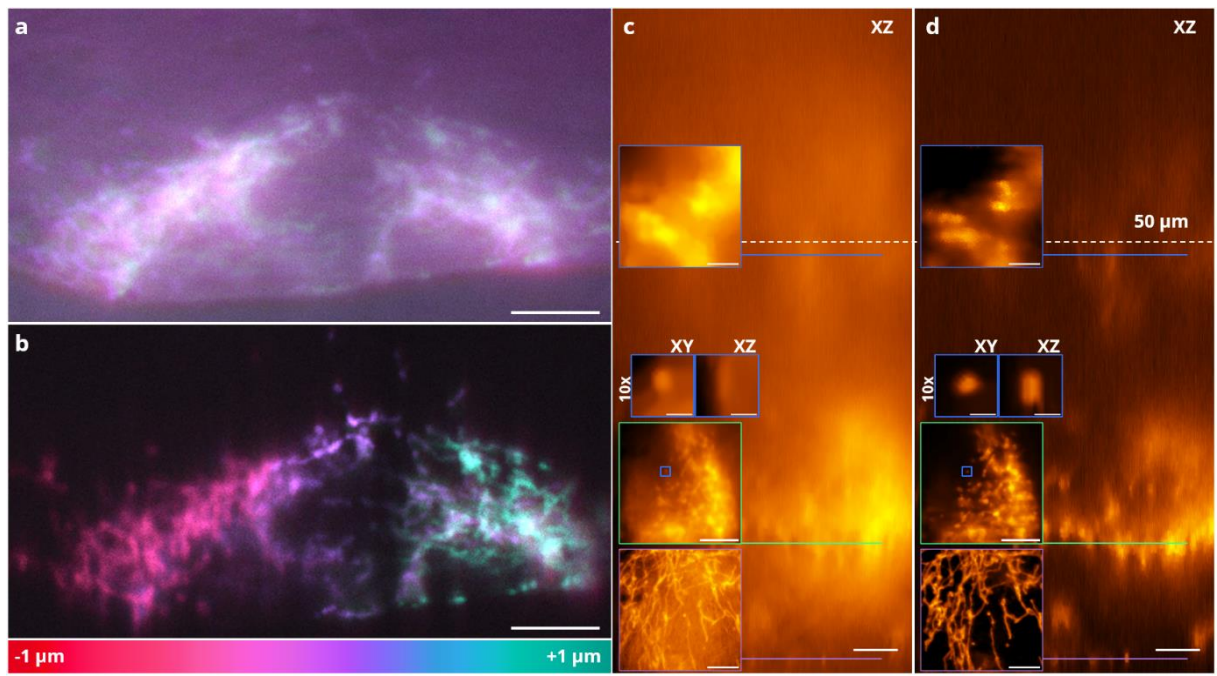

*Figure 4: Engineered human heart tissue imaging. (a) Multifocus imaging spanning 2 µm of a cell with labelled mitochondria (TOM20), several micrometres inside the tissue. The cell is displayed as a color-coded maximum intensity projection. (b) The same cell imaged with SOLIS. (c,d) A z-stack in side-view. Shown is a single panel of the multifocus imager in widefield and SOLIS mode. Individual mitochondria are discernible up to 30 µm deep into the tissue. Beyond 50 µm depth, only larger agglomerates are discernible. The inlays show denoised XY sections at 2 µm, 15 µm, and 48 µm depth with a mitochondria-derived vesicle highlighted in the 10x view. Scale bars are 5 µm and 500 nm in the 10x views.*

## Discussion

Multifocus microscopy encompasses an arsenal of techniques to multiplex a 3D sample onto separate 2D locations on a camera in a single frame. Thus acquired images generally lack optical sectioning, which is elemental for volumetric imaging of dense structures and poses a limit of multifocus microscopy. To alleviate this drawback, we demonstrated SOLIS, an approach to record entire optically sectioned volumes in single camera exposures. This is possible by conjugating a swept illumination plane with the light-sheet read-out mode of rolling shutter cameras. In effect, this combination realises a plane-scan version of confocal theta microscopy [26], [27]. Thus, SOLIS acquires an optical sectioning performance comparable to spinning-disk microscopy yet at much higher volumetric frame-rate due to its multi-plane characteristic. A further advantage of SOLIS over spinning-disk microscopy is the reduced crosstalk as SOLIS shares more characteristics with line-scanning as compared to point-cloud scanning with Nipkow disks.

We demonstrated 35 volumes per second, which should not be seen as an upper limit. At higher speeds, it is important to ensure good synchronisation between illumination and read-out scan, which requires a high linearity in the galvanometer scanner. Recently, a scan multiplier approach was presented [28] that permits generation of such highly linear scans far beyond the inertia limit. When combined with latest camera technology (Hammamatsu Fusion or Photometric's Kinetix) that support kilohertz frame rates, one could achieve scan rates an order of magnitude faster than demonstrated in our setup.

If optical sectioning is not necessary to be achieved in a single camera frame, alternative optical sectioning approaches exist for multifocus microscopy. Super-resolution optical fluctuation imaging (SOFI) [13], [15] and structured illumination microscopy (SIM) [29] have been combined with multifocus microscopy. Both approaches require several volumes to be recorded sequentially, which are then processed into a single sectioned volume. They are thus not truly single shot techniques but do promise resolution gains both axially and laterally. Pure single-shot sectioning could be realised through a variant of optical sectioning SIM that uses polarization-coding (picoSIM) [30]. Here, retained fluorescence polarisation enables encoding of multiple frames in a single camera exposure. This idea was conceptually combined with multifocus optics [31] and shown in a proof-of-concept study but never realised in practice. Note that this approach is strongly limited by fluorescent labels, which need to exhibit highest possible fluorescence anisotropy.

Apart from volumetric imaging speed, SOLIS permits high light efficiency. As each illuminated plane is recorded, SOLIS compares favourably to confocal techniques and is in fact closer related to light-sheet systems with axial sweeping (ASLM) [32]. Both SOLIS and ASLM gain axial resolution by trading some light-efficiency due to light-rejection during the rolling shutter read-out. In both techniques, this is the key ingredient for optical sectioning and increased axial resolution. As SOLIS is a single objective technique, it does not share the space constraints of two-objective microscopes like ASLM and thus can utilise highest numerical apertures. We calculate the overall collection efficiency to be up to 53% higher (1.1 NA versus 1.5 NA) in favour of SOLIS. When factoring in light-loss incurred by multi-plane optics (estimates are 10%-20% [33]), the overall increase in photons captured is still more than 23%.

Multifocus microscopy with higher light efficiency can be realised through reflective pinhole- or slit-cascades in an intermediate image plane [16], [17]. This has some advantages over prism or grating based multiplexing. If only a small number of planes is required, slit-cascades can rival light-sheet microscopy in terms of light-efficiency. However, pinhole cascades use separate detectors for each plane and, as pinhole cascades are effectively point-scanning, they are limited in their maximally achievable framerate. Even with faster scanner and better detectors, a point-scanner is ultimately limited by the sample's fluorescence lifetime, which puts a lower bound on the pixel dwell time to achieve a usable signal-to-noise ratio.

Slit cascades do not suffer this limitation in practical scenarios, but they do require individual aberration correction for each plane, which hinders efficient scaling. So far, only a 3x1 slit cascade has been demonstrated. Furthermore, slit cascades are challenging to realise with Nyquist sampling along the axial direction in high NA systems due to space limitations in the intermediate image plane. In contrast, SOLIS could be scaled with aberration-corrected multifocus gratings up to 25 planes [20] at Nyquist sampled inter-plane distances. Such implementations benefit from cameras with multi-line rolling shutters, which are already commercially available. Currently, manufacturers offer cameras with two rolling shutters (pco.edge, pco) and are expected to developed cameras with even more parallel readout shutters (e.g. expected from Kinetix v2, Photometrics).

In summary, we introduced a scalable multifocus microscopy method dubbed SOLIS that incorporates optical sectioning and axial superresolution capabilities. We derived the theoretical framework, which was verified in simulations, and constructed a prototype system with a 3x1 beam splitter cascade at its core. We imaged BPAE cells at 35 volumes per second and recorded the distribution of mitochondria and mitochondria derived vesicles in 2 µm thick instant-volumes up to 30 µm deep into uncleared engineered human heart tissue. We demonstrated axial resolution gains of over 200 nm in case of SOLIS over widefield microscopy.

## Methods

### Theoretical estimation of SOLIS resolution

Let us denote the light-sheet illumination as $h_L(x)$ and the detection point spread function as $h_D(x)$. An image $i(x)$ formed be a regular widefield microscope with unit magnification of a fluorescent sample $s(x)$ is thus described by

$$i(x) = h_D(x) \otimes [h_L(x) \times s(x)]$$

*Equation 1*

Uniformly moving the light-sheet as $h_L(x - m)$ during a global exposure cycle of the camera integrates over the sweeping variable $m$ and thus eliminates the light-sheet from the equation up to a constant (omitted). The imaging model is that of widefield microscopy

$$i(x) = h_D(x) \otimes \left[ \int h_L(x - m) \times s(x) \, dm \right]$$

$$i(x) = h_D(x) \otimes \left[ s(x) \times \int h_L(x - m) dm \right]$$

$$i(x) = h_D(x) \otimes s(x)$$

*Equation 2a-c*

In case of a rolling shutter $r(x)$ that is synchronised to the light-sheet, the detection point spread function becomes dependent on the sweeping variable.

$$i(x) = \left[\int h_D(x) \times r(x - m) \, dm\right] \otimes \left[\int h_L(x - m) \times s(x) \, dm\right]$$

*Equation 3*

If the rolling shutter is narrow, it can be approximated with a delta pulse $\delta(x)$ in sweep direction. Reversing the order of the two convolution integrals and using the delta pulse convolution shift theorem, the imaging equation becomes

$$i(x) = \left[\int h_D(x) \times \delta(x - m) \, dm\right] \otimes \left[\int h_L(x - m) \times s(x) \, dm\right]$$

$$i(x) = \iint [h_D(x') \times \delta(x' - m)] \times [h_L(x' - m) \times s(x' - x)] \, dm \, dx'$$

$$i(x) = \iint h_D(x') \times [\, \delta(x' - m) \times h_L(x' - m)] \times s(x' - x)] \, dm \, dx'$$

$$i(x) = \int h_D(x') \times h_L(x') \times s(x' - x) \, dx'$$

$$i(x) = [h_D(x) \times h_L(x)] \otimes s(x)$$

*Equation 4a-e*

The effective point spread function thus consists of the multiplication of light-sheet and detection point spread function and the overall optical transfer function is the convolution of the respective transfer functions. The resolution limit is hence the sum of the constituting transfer function limits. An oblique light-sheet spanning half of the illumination NA therefore provides the same axial resolution limit as provided by structured illumination microscopy, roughly twice over the axial resolution limit of widefield microscopy [34]–[36]. In the general case of a wider rolling shutter, or thicker light-sheet Equation 3 governs the image formation, and the expected resolution gain becomes smaller.

## Simulations

A cube with side length 6.4 µm was simulated in MATLAB at 100 nm voxel size, dotted with randomly distributed point emitters. We employed Fiji's [37] PSF generator plugin [38] to generate PSFs with the Gibson & Lanni model (Immersion RI = 1.4, Sample RI = 1.38, NA = 1.35, WL = 550 nm), which were convolved with the point emitters for widefield imaging. In case of SOLIS, a light-sheet was moved pixel-wise through the volume and multiplied with point emitters before convolution and rolling-shutter application. Light sheets were created using PSF from the aforementioned PSF generator plugin, followed by averaging of the PSFs along X as to generate sheets. In case of oblique illumination, the sheet was tilted. Light-sheet NAs were chosen such that the PSFs' axial extent covers the simulated volume, which equates to 0.483 NA for an on-axis light-sheet and 0.377 NA for a maximally oblique light-sheet. We used axial Abbe resolution as metric for light-sheet length. The simulation results are presented in Figure 2a-c. For panels d-g, a single point emitter was simulated with otherwise unchanged parameters. OTFs were calculated from PSFs with Fiji's fast FFT plugin.

## Optical system

SOLIS' light path is depicted in Figure 1a. A 25 mm focal length cylindrical lens (Cyl; 68160, Edmund Optics) shapes a collimated 488 nm laser beam (Fisba READYbeam) into a light-sheet, which is relayed by 39 mm and 70 mm focal length scan lenses (SL1 and SL; LSM03-VIS and CLS-SL, Thorlabs) over a galvanometric mirror (GM; GVS211, Thorlabs) into a conjugate image plane and over a 200 mm focal length tube lens (TL1, TTL200, Thorlabs), a dichroic mirror (DM; Di03-R405/488/532/635, Semrock), and an objective into the nominal sample plane. A 0.95NA 40x dry objective or a 1.35NA 100x silicone immersion objective (both Nikon) were used. Decentering the light path before the GM allows inclination of the light-sheet. Fluorescence is collected episcopically and relayed through a 525/45 emission filter (FF01-525/45, Semrock), a 200 mm focal length tube lens (TL2, TTL200, Thorlabs), and

3x1 beam splitter cascade onto an sCMOS camera (BSI Express, Photometrics) with around 10 mm optical path difference between the focal planes: the beam splitter cascade consists of a 30:70 and a 50:50 non-polarizing beam splitter (BS052 and BS004, Thorlabs) and a right-angle prism (PS914L-A, Thorlabs) to relay the transmitted light onto the camera with approximately the same optical path difference as between the reflected paths of the beam splitters. A small adjustable distance (0-2mm) between the beam splitters and the right-angle prism adds to the path length differences and allows for fine-tuning. As the camera chip has a side length of 13.3 mm, the shortest and longest beam paths require about 2.5° inclination of the first beam splitter and the right-angle prism in opposite directions, which results in an additional path difference of about 0.5 mm in the same direction for both outer focal planes. Conventional multifocus widefield imaging was realized by scanning the light-sheet once during a global exposure of the camera, while SOLIS imaging synchronized the light-sheet scan with the line-scan mode of the camera's rolling shutter using a DAQ board (PCIe-6738, NI). Note that the programmable line-scan mode is a feature of latest sCMOS cameras but can be emulated in a conventional rolling-shutter camera by setting the exposure time close to the line-time of the sensor. We generally achieved a good trade-off between speed, light-efficiency, and sectioning capability with a scanning linewidth of 3 pixels.

BPAE cell imaging and analysis

Imaging experiments in Figure 3a and b were performed on commercially available fixed bovine pulmonary artery endothelial cells labelled with Alexa Fluor™ 488 phalloidin to stain actin (F36924, Thermo Fisher Scientific). We used a 0.95NA 40x dry objective, which resulted in a plane separation of -5.6 µm and +5.4 µm with a frame rate of 8 volumes per second. The same cells were also imaged with a 1.35NA 100x silicone objective at 35 volumes per second with a separation of 1 µm between each plane while moving the stage to emulate a fast-moving sample. This is shown in Visualisation 1.

## Engineered human heart tissue (EHT) preparation

The human induced pluripotent stem cell (hiPSC) line (UKEi003-C) was differentiated into cardiomyocytes using a 2D monolayer protocol. This cell line was kindly provided by the Institute of Experimental Pharmacology and Toxicology, University Medical Center Hamburg-Eppendorf and is registered at the European Human Pluripotent Stem Cell Registry (hPSCreg). EHT was produced as previously described (Breckwoldt et al. 2017) with 106 hiPSC-derived cardiomyocytes embedded in fibrin hydrogel. After more than 21 days in culture, the beating EHT was fixed in 4% PFA at 4°C overnight. Immunofluorescent staining of mitochondria in the fixed EHT was performed with anti-TOM20 antibody (Santa Cruz) and Alexa Fluor® 488 anti-rabbit antibody.

## Engineered human heart tissue imaging and analysis

EHTs were imaged using a 1.35NA 100x silicone objective with multi-focus z-span of 2 µm. To create z color-coded images as shown in Figure 4a and b, we inserted additional frames by cubic interpolation between the recorded nominal planes before applying Fiji's color-coded maximum intensity projection function. The images displayed in the inlays of Figure 4c and d where denoised using Fiji [39]. Line profiles through MDVs were fitted with Gaussian functions and standard deviations were converted to FWHM using the required conversion factor of $2\sqrt{2\ln(2)} \approx 2.355$.

## Bead imaging and analysis

The 200 nm Tetraspeck fluorescent beads (T7280, Thermo Fisher Scientific) displayed in Figure 3c were imaged with a 1.35NA 100x silicone objective. Line profiles were fitted with Gaussian functions using the curve fitting plugin of Fiji. The found standard deviations were converted to FWHM using the conversion factor $2\sqrt{2\ln(2)} \approx 2.355$ and reported as un-corrected resolution. To remove bead size as a factor from the measured FWHM values, we simulated widefield imaging of a 200 nm diameter spherical shell to approximate the used beads. Line profiles through this image were fitted with Gaussian functions and the corresponding FWHM divided by the FWHM of the PSF of a 100 nm spherical shell to obtain a correction factor $c = 0.8417$. Using this factor, we can correct for the bigger real bead size and compare the simulation results stated in Table 1 with the measurements displayed in Figure 3.

## Data availability

The datasets generated during and/or analysed during the current study are available in the *DataverseNO* repository: dataverse.no/dataset.xhtml?persistentId=doi:XXX

## Acknowledgements

The authors would like to thank Digital Life Norway and Aurora Outstanding for support of this work.

## Author contributions

FS conceived the project, derived the theoretical framework, built the microscope, performed simulations and imaging, analyzed the data, and wrote the manuscript. FS and DHH wrote control software for the microscope. ÅBB cultured and prepared EHTs. All authors commented on the manuscript.

## Funding

This work was funded by the Norwegian Research Council (project no. 314546) and the EU's Horizon 2020 program (project no. 964800).

## Disclosures

UiT The Arctic University of Norway has applied for patent on "Volumetric Imaging" with Florian Ströhl as inventor (pending UK patent application number 2111782.5). The patent covers the technique for optical sectioning of multifocus microscopes.

# Supplementary Data

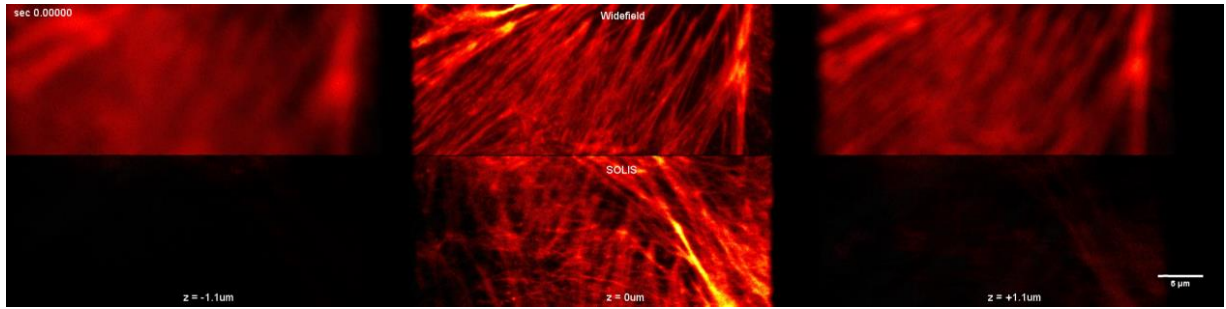

*Visualization 1: BPAE cells imaged at 35 Hz with three simultaneous planes, while the stage was moved at 35 µm/s. A 1.35 NA 100x objective was used to realise a plane separation of 1.1 µm. Top row shows conventional multifocus microscopy, the bottom row shows SOLIS. Note that widefield and SOLIS do not show the identical cells, as collection was performed sequentially.*

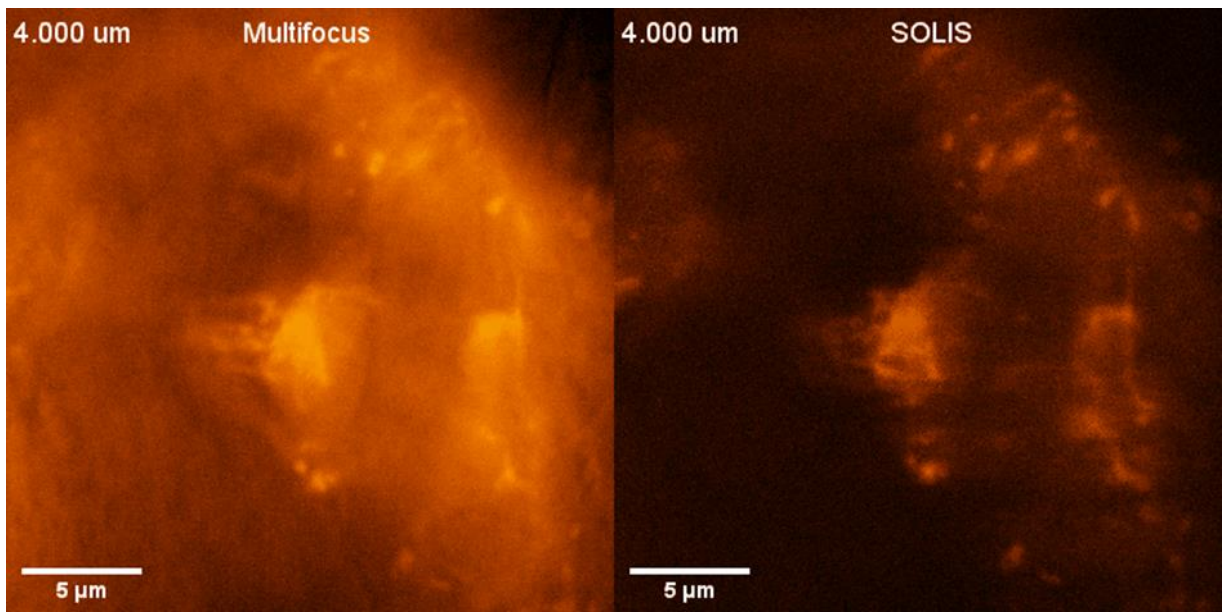

*Visualization 2: Comparison of conventional multifocus microscopy and SOLIS on cardiomyocytes within engineered human heart tissue. Shown are mitochondria labelled with TOM20-Alexa488, around 5 µm deep inside the uncleared tissue. The 2 µm thick image volume was captured in a single exposure and the fly-through generated with cubic interpolation between the collected z-planes*

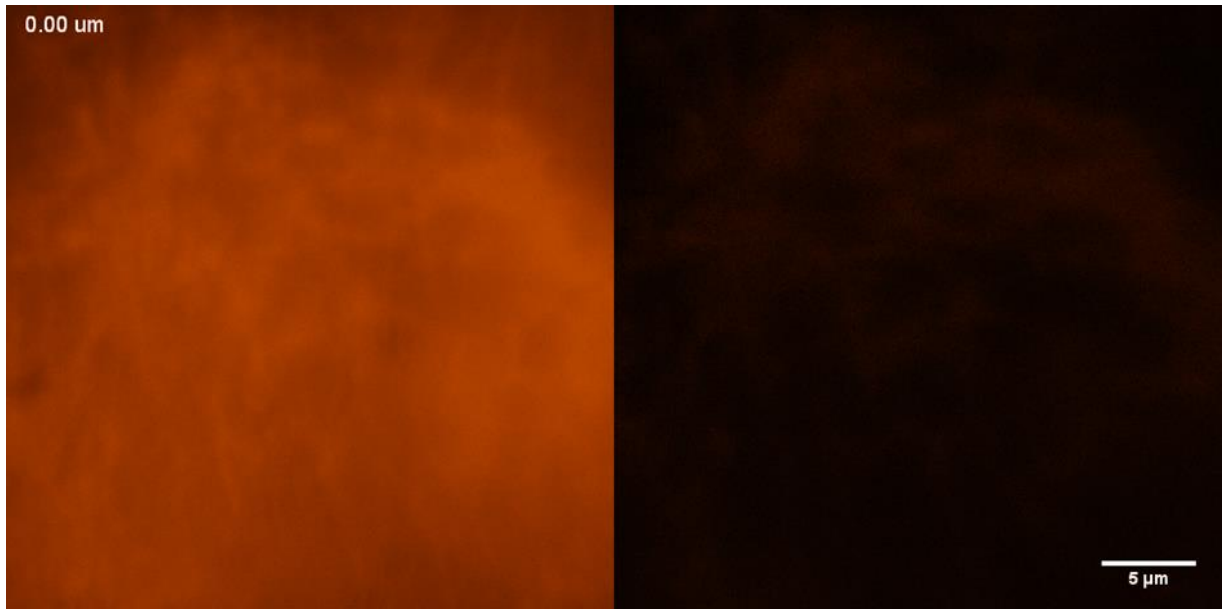

*Visualization 3: Comparison of the central slice of conventional multifocus microscopy (left) and SOLIS (right) on cardiomyocytes within engineered human heart tissue. Shown are mitochondria labelled with TOM20-Alexa488, while scanning to a depth of around 38 µm deep into the uncleared tissue.*